\documentclass[prb,preprint,showpacs]{revtex4}

\usepackage{t1enc}
\usepackage[final]{graphicx}
\usepackage{graphicx}
\usepackage{epsfig}

\begin{document}

\title{Exponential equilibration by slow cooling in the planar random-anisotropy magnet: experiments and simulations}
 
\date{\today}

\author{Donald R. Taylor}
\email{taylordr@physics.queensu.ca}
\affiliation{Department of Physics, Engineering Physics, and Astronomy, Queen's University, Kingston, ON Canada K7L 3N6}

\author{Orlando V. Billoni}
\email{billoni@famaf.unc.edu.ar}
\affiliation{Instituto de F\'{\i}sica de la Facultad de Matem\'atica, Astronom\'\i a y F\'\i sica (IFFAMAF-CONICET),
         Universidad Nacional de C\'ordoba,
         Ciudad Universitaria, 5000 C\'ordoba, Argentina}

\author{Sergio A. Cannas}
\affiliation{Instituto de F\'{\i}sica de la Facultad de Matem\'atica, Astronom\'\i a y F\'\i sica (IFFAMAF-CONICET),
         Universidad Nacional de C\'ordoba,
         Ciudad Universitaria, 5000 C\'ordoba, Argentina}

\author{Francisco A. Tamarit}
\affiliation{Instituto de F\'{\i}sica de la Facultad de Matem\'atica, Astronom\'\i a y F\'\i sica (IFFAMAF-CONICET),
         Universidad Nacional de C\'ordoba,
         Ciudad Universitaria, 5000 C\'ordoba, Argentina}

\begin{abstract}

Neutron measurements of the equilibration of the staggered magnetization in DyAs$_{0.35}$V$_{0.65}$O$_{4}$ are compared with Monte Carlo simulations of spin dynamics in a planar random-anisotropy magnet. The simulation results are in agreement with striking observed relaxation phenomena: when cooled rapidly to a low temperature no magnetic ordering is observed, but when cooled in small steps an ordered magnetic moment appears which is found to equilibrate exponentially with time at temperatures through and below the transition temperature. In contrast to the freezing of spins in other random systems, the time scale of the relaxation in this system does not increase significantly even at the lowest temperatures. 
\end{abstract}


\maketitle

\section*{Introduction}

The presence of disorder in magnetic systems generally leads to characteristic slow (`glassy') relaxation of the magnetization at low temperatures. In typical spin glasses there is no sharp magnetic ordering transition and no spin equilibration on laboratory time scales\cite{Nordblad}. Models for analysis and simulations of disordered magnets generally include random exchange and random uniaxial anisotropy, which are believed to be the most relevant mechanisms for spin glass properties\cite{Sellmyer}. There is an extensive literature on the effects of both random exchange and random uniaxial anisotropy, with most emphasis on random exchange interactions which are certainly important in the commonly studied metal alloy systems. However some recent experiments \cite{Bert} and simulations\cite{BiCaTa2005} have demonstrated the important role of random anisotropy on the spin freezing and relaxation processes in these systems. It should be possible to study the effects of random anistropy separately from random exchange by choosing systems where the latter should not be significant. This can be achieved in crystalline samples by not diluting or mixing the magnetic ions but instead by partial substitution of neighboring ions. Some experiments in such a system, DyAs$_{x}$V$_{1-x}$O$_{4}$, were reported previously\cite{{Taylor1},{Taylor2}} and are further discussed here. Measurements of the magnetization relaxation in this material, which is believed to be a  random planar anisotropy magnet (RPAM) showed surprising behavior\cite{Taylor2}: for example, in contrast to other random magnetic systems the magnetization relaxation times did not become very long at lower temperatures but remained roughly constant or even decreased. To address the question as to whether this property was a fundamental characteristic of RPAM systems or whether it had some other origin we have carried out Monte Carlo simulations on the RPAM model and compared the results with experiments. Good agreement was found between the simulations and the observed relaxation behavior, confirming that it is characteristic of the RPAM system. 

\section*{Experiments}

First the sample and the experiments will be briefly described: further information can be found in Refs. \onlinecite{{Taylor1},{Taylor2}}. In DyAs$_{0.35}$V$_{0.65}$O$_{4}$ the Dy-Dy interactions are expected to drive an antiferromagnetic ordering transition at a temperature of a few K. The As/V substitutions generate uniaxial random anisotropy at each Dy site without significantly altering the Dy-Dy interactions. The Dy spins lie in the basal plane of the tetragonal structure and order parallel to one or the other basal plane axes depending on the local anisotropy. In neutron experiments carried out at the Canadian Neutron Beam Centre, Chalk River, the growth of the antiferromagnetic (100) peak was studied as the sample was cooled below 1.6 K. No ordering field is applied in these experiments: the peak intensity grows in response to temperature reductions giving a direct measure of the staggered magnetization. It should be noted that the study of magnetic peaks by neutron scattering has rarely been an effective technique in investigations of spin glasses and other disordered magnets. In many samples the severe chemical disorder destroys the lattice periodicity; in others where there may be a well-defined lattice the random magnetic interactions lead to a disordered short-range spin freezing structure that gives only a very broad magnetic diffraction peak. The present system is very favorable for diffraction studies since the lattice is only slightly disordered, and it is known that planar interactions are able to drive a magnetic transition with quasi-long range order even in the presence of strong random anisotropy\cite{Fisch}. As shown previously\cite{{Taylor1},{Taylor2}} the neutron experiments do show a clear but slightly rounded magnetic transition at $T_c$=1.6 K, with a magnetic peak that is quite narrow but not resolution-limited, implying correlation lengths of about 40 nm.

\begin{figure}[t]
\includegraphics*[width=11cm,angle=0]{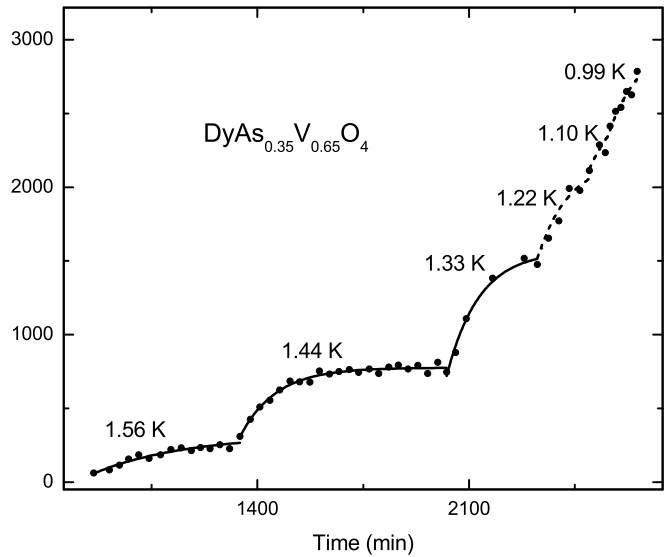}
\caption{\label{fig1}Growth of (100) magnetic peak intensity while cooling in small steps. Solid lines are fits to exponential equilibration; dashed lines are guides to the eye.} 
\end{figure}

Our experiments showed that when the sample was cooled rapidly to a temperature well below 1.6 K there was no detectable (100) peak, and hence no magnetic ordering, even after waiting 24 hours. However, when the temperature was reduced in small steps, typically 0.1 K, and held constant for a few hours at each temperature while the (100) peak was scanned repeatedly, the magnetic peak grew and reached a stable intensity. To achieve temperatures below 1 K a $^{3}$He cryostat was used but since its cycle time was limited to $\sim30$ hours the time available at each temperature was significantly restricted. As a result the experimental statistics are relatively poor, but the main results are nevertheless clear. The growth of the magnetic peak intensity as a function of time is shown in Fig. \ref{fig1} along with exponential fits. The long set of data at 1.44 K shows exponential equilibration most convincingly. The equilibration times obtained for the fits were as follows: 307 $\pm$ 135 min at 1.56 K; 118 $\pm$ 10 min at 1.44 K; 107 $\pm$ 23 min at 1.33 K. At the three lowest temperatures where the data were too sparse to allow unrestricted fitting, the dashed lines are exponential fits with the relaxation time set equal to 107 min. They are guides to the eye but provide evidence that the relaxation times are not lengthening rapidly at these low temperatures, in contrast to the pronounced freezing of the spin dynamics found in most disordered magnets.

The above results can be put into meaningful context by comparison with results for identical experiments carried out on a sample with reduced As substitution, DyAs$_{0.17}$V$_{0.83}$O$_{4}$. In this sample the structural disorder was not large enough to inhibit the tetragonal to orthorhombic transition that occurs in pure DyVO$_{4}$ at 14 K, but instead only suppressed the transition temperature to about 8 K\cite{Taylor3}. As a result the random crystal fields are overwhelmed by the uniform orthorhombic distortion and we do not expect a random anisotropy effect on the magnetic transition, although some disorder still exists. In this sample a strong (100) magnetic peak appeared signalling magnetic ordering at 2.3 K. In contrast to the behavior of the sample with 35\% As concentration which remains tetragonal on average with random anisotropy axes, the (100) peak was resolution limited indicating long range order, and its growth showed no observable time dependence. Although these neutron experiments cannot access the actual relaxation times, we expect that they are characteristic of a pure system, that is of order a spin-flip time $h/J$ extended by critical slowing-down.

\section*{RPAM simulations}

The system is described by the following classical Heisenberg Hamiltonian:

\begin{equation}
H=-J \sum_{<i,j>} \vec{S}_{i} \cdot \vec{S}_{j} - D \sum_{i} (\vec{n}_i
\cdot \vec{S}_i)^2 
\end{equation}
where $D$ and $J$ are the anisotropy and the exchange couplings respectively. The spin variable $\vec{S}_i$
is a three component unit vector associated with the $i$--th node of a cubic lattice and the first sum runs over all nearest-neighbor pairs of spins. $\vec{n}_i$ is a unit
random vector that defines the local easy axis direction of the anisotropy at site $i$.  These easy axes are quenched variables chosen from a given distribution: the random planar anisotropy magnet (RPAM) model is defined by choosing an isotropic distribution on a circle in the $x-y$ plane. 

The simulations were performed in a system of
$N=L^3$ spins using a Monte Carlo Metropolis algorithm with periodic boundary conditions. In the case of the RPAM we set $J>0$ when the neighboring spins are in the $x-y$ plane and $J<0$ in the other case, so when $D=0$ we have a planar
antiferromagnet.

\begin{figure}[t]
\includegraphics*[width=9.5cm,angle=-90]{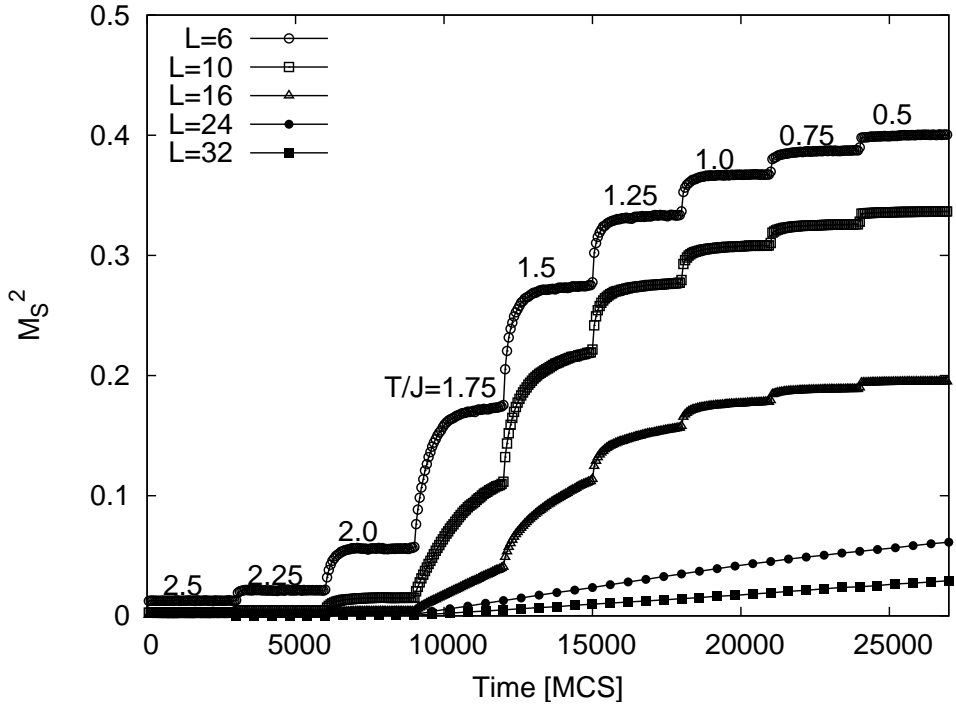}
\caption{\label{fig2}Relaxation of the staggered magnetization for different system sizes (from top to bottom L=6, 10, 16, 24, and 32). The anisotropy to exchange ratio is $D/J=7$.}
\end{figure}

In order to  follow the cooling protocol outlined in Ref. \onlinecite{Taylor1},  we  started every run at a  temperature above the ordering temperature 
(around $T/J=1.8$) and then  reduced it in uniform steps. Immediately after each step the relaxation
of the root mean square staggered magnetization is measured during a time period larger than the longest observed relaxation time. Times are expressed in Monte Carlo steps (MCS) where a MCS is defined as a complete cycle of N spin update trials. The results are shown in Fig. \ref{fig2} for $D/J=7$ and different system
sizes; each curve was averaged over $10^2-10^3$ samples.   For system sizes  $L=24$ and $L=32$  only temperatures larger than $T/J=1.75$  are shown. It can be seen that the relaxation times show qualitatively the same behavior as in Fig. \ref{fig1}, in particular the relaxation times do not increase significantly at low temperatures. Fig. \ref{fig2} also shows that the staggered magnetization takes  lower values in the whole temperature range when the system
size increases, suggesting, as expected \cite{Fisch}, that the ordered state has zero staggered magnetization in the  thermodynamic limit. In contrast, the magnetic peak observed experimentally is not much smaller than that for the conventional antiferromagnet DyAs$_{0.17}$V$_{0.83}$O$_{4}$. This disagreement could arise from a small difference between the interactions in the real system from those assumed in the model, for example the distribution of anisotropy directions may not be uniform in the basal plane. 

\begin{figure}[t]
\includegraphics*[width=9.5cm,angle=-90]{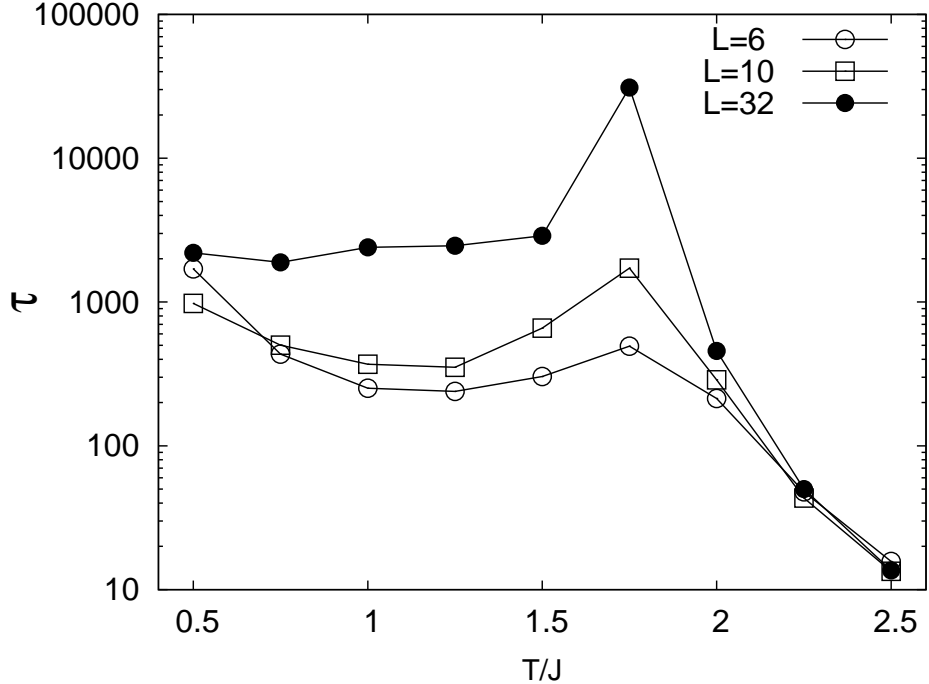}
\caption{\label{fig3}Relaxation time vs. temperature for the RPAM model (obtained from fittings of curves shown in Fig. \ref{fig2}).}
\end{figure}

The relaxation curves shown in Fig. \ref{fig2} are well fitted by exponential functions and we show the relaxation times $\tau$ in Fig. \ref{fig3} as a function of temperature.  A distinct peak in the relaxation times can be observed at $T/J = 1.75$  for every system size; the height of the peak increases with the system size. This behavior strongly resembles the critical slowing down properties for a system undergoing a conventional second order phase transition. Below $T_c$ the relaxation times drop to a plateau where they remain relatively unchanged down to low temperatures (especially for large $L$). Our experiments can only provide data below $T_c$ for comparison but they show the same qualitative behavior as the simulations below $T_c$, with $\tau$ initially decreasing and then remaining roughly constant. 

\begin{figure}[b]
\includegraphics*[width=9cm,angle=0]{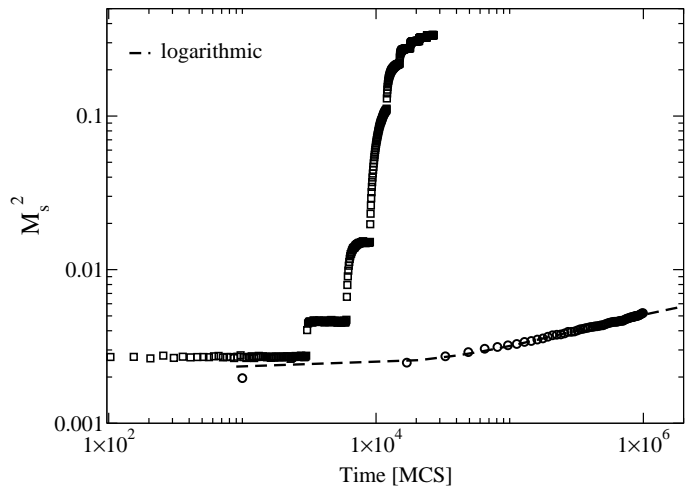}
\caption{\label{fig4}Relaxation of $M_S^2$ after quenching to $T/J=0.5$ (L=10) together with logarithmic fit. For comparison the relaxation after slow cooling down to the same temperature for L=10 as shown in Fig. \ref{fig2} is also included.}
\end{figure}

In contrast, when we quenched the system to $T<T_c$ in our simulations the staggered magnetization was found to grow on a time scale much greater than those observed for slow cooling. Fig. \ref{fig4} illustrates this behavior for a quench to $T/J=0.5$. The slow relaxation can be fitted by a logarithmic function as shown, although a stretched exponential also gives an adequate fit. 

\begin{figure}[t]
\includegraphics*[width=9.5cm,angle=-90]{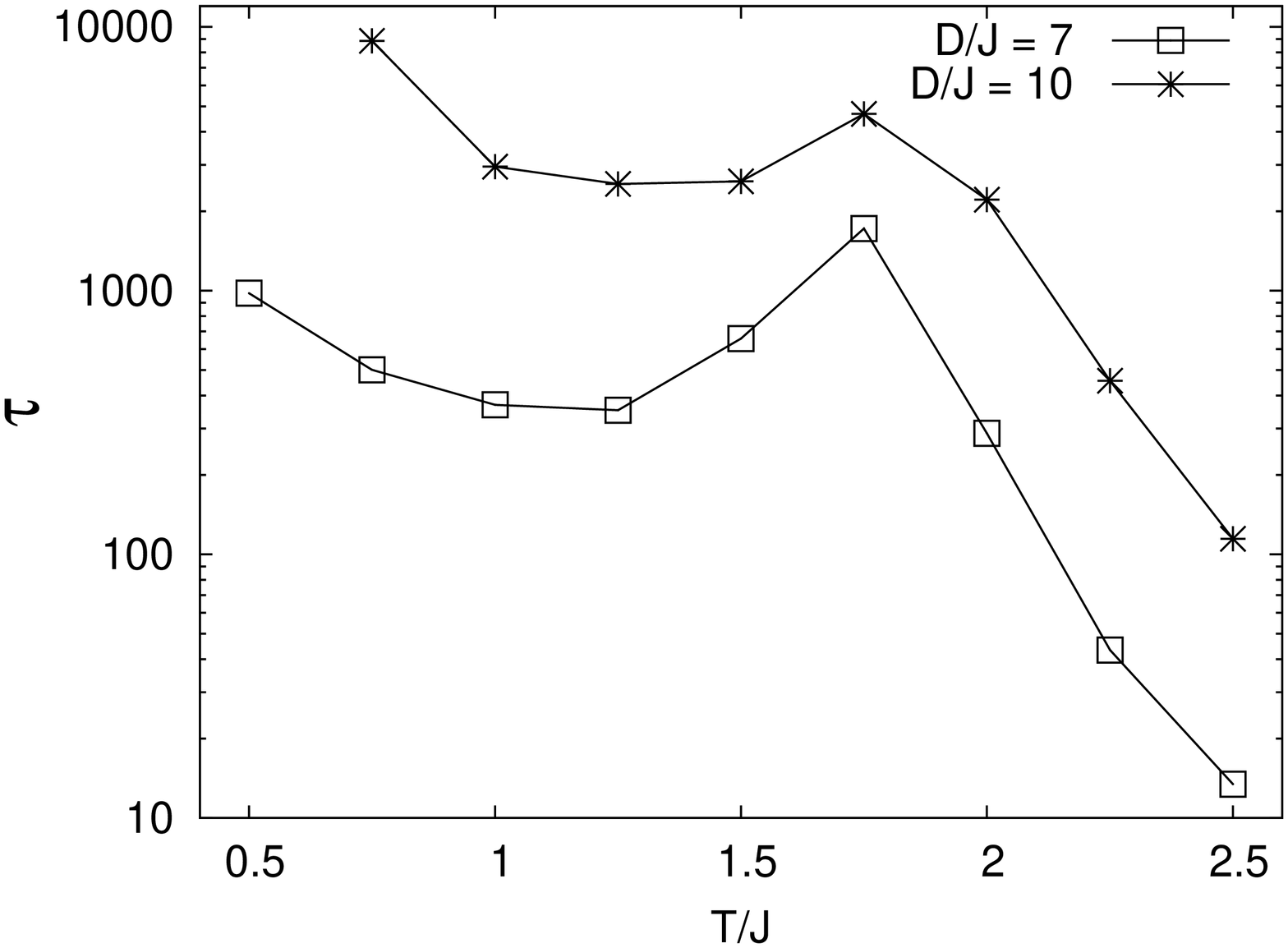}
\caption{\label{fig5}Relaxation times in the RPAM model, for L=10 and different anisotropy strengths.}
\end{figure}

The value of $D/J$ appropriate to our sample is unknown, but we expect it to be large because in DyVO$_4$ the ion-lattice interactions are strong and induce a crystallographic phase transition at a temperature much higher than the magnetic transition. We determined relaxation times in simulations with $D/J$ values of 7 and 10 and show the results in Fig. \ref{fig5}. While the critical temperature and temperature dependence seem to be independent of the anisotropy, the most remarkable result is the rapid, perhaps exponential, increase of the low temperature relaxation time with $D/J$. This suggests an Arrhenius mechanism and can be understood by the increase of the local barriers due to the anisotropy. A comparison of MCS and experimental time scales confirms that a large $D/J$ value is appropriate, but we have not attempted to match the time scales by repeating the simulations with larger $D/J$ values because the model is too simple in some respects, for example, the magnetic interactions are assumed nearest neighbor only, and only a single value of $D$ is used. 

\section*{Discussion}

The relaxation properties observed here are quite different from those of other spin systems. For example, if non-interacting spins are cooled in an applied field, exponential equilibration occurs with a characteristic time constant, and if the system is quenched and observed on a much shorter time scale there would indeed be no observed relaxation. The key difference is that in that system, unlike ours, the relaxation time does not depend on the cooling rate. Moreover, since relaxation depends on thermal processes, it slows dramatically at lower temperatures in contrast to the temperature-independent behavior that we observe under slow cooling. If we compare our system with typical spin glasses, in both cases the spins freeze and are unable to order after rapid cooling, but our system differs in being able to equilibrate fully even at the lowest temperatures when cooled slowly. In some magnetic systems the observation of temperature-independent relaxation has been explained by a quantum tunnelling mechanism\cite{Tejada}. However those observations are concerned with the response of the system to a magnetic field, whereas in our case we are dealing with a purely thermal response, namely the growth of the staggered magnetization following a reduction in temperature. This requires energy exchange between the spin system and a thermal reservoir, a scenario quite different from that in which quantum tunnelling operates. 

In the hope of clarifying the relaxation mechanism in our system and determining if it is unique to the RPAM system we have carried out further simulations using the same cooling protocol on the 3d RAM system and the pure Heisenberg model. Full results will be reported later\cite{BiCaTa2008} but preliminary results show some significant similarities. Like the RPAM system, the 3d RAM system shows glassy dynamics when cooled rapidly but equilibrates exponentially when cooled slowly. The relaxation times do not show a well-defined peak, consistent with the expected lack of long-range order for the 3d RAM\cite{{Sellmyer},{It2003}} but reach a plateau with little variation with temperature at lower temperatures. In the pure Heisenberg system the magnetization does not show glassy dynamics but the relaxation times show a temperature dependence that is qualitatively similar to the RPAM system, with a well-defined peak corresponding to critical slowing down and a plateau at lower temperatures. The actual relaxation times are, of course, many orders of magnitude shorter. The surprising inference is that random anisotropy does not qualitatively alter the relaxation dynamics if the system is slowly cooled, although the relaxation times are increased by orders of magnitude. Thus it appears that the free energy barriers due to random anisotropy in this system are not too high and under slow cooling the system is able to reach states that are close to the the equilibrium state. It should be noted that a distribution of energy barrier heights is expected in the experimental system because of the random atomic substitutions. Although in the simulations a single D value is assumed, the anisotropy energy barriers encountered by coupled spins should also have a random height distribution because each block of spins will have a different combination of favorable and unfavorable axes. Thus it is unexpected that the relaxation on slow cooling is found to follow a simple exponential time dependence in both the experiments and the simulations. In the case of rapid quenching the system 
evidently gets stuck in a local minimum of the rough free energy 
landscape and relaxation is thereby inhibited.

The collaboration of W. J. L. Buyers and J. T. Love in the experimental work is gratefully acknowledged. Research support was provided to D. R. Taylor by NSERC (Canada). 
Partial support was also provided by grants from CONICET, SECyT-UNC and FONCyT grant PICT-2005 33305 (Argentina).

\bibliographystyle{prsty}

\begin{thebibliography}{100}

\bibitem{Nordblad} P. Nordblad and P Svedlindh, in \textit {Spin Glasses and Random Fields},
edited by A. P. Young (World Scientific, Singapore, 1998), p. 1.
\bibitem{Sellmyer} D. J. Sellmyer and M. J. O'Shea, in \textit {Recent Progress in Random Magnets}, edited by D H. Ryan (World Scientific, Singapore, 1992), p. 71.
\bibitem{Bert} F. Bert, V. Dupuis, E. Vincent, J. Hammann, J.-P. Bouchaud, Phys. Rev. Lett. \textbf{92}, 167203 (2004).
\bibitem{BiCaTa2005} O. V. Billoni, S. A. Cannas, F. A. Tamarit, Phys. Rev. B \textbf{72}, 104407 (2005).
\bibitem{Taylor1} D. R. Taylor, J. T. Love, E. M. Sheridan, W. J. L. Buyers, J. Magn. Magn. Mater. \textbf{310}, 1473 (2007).
\bibitem{Taylor2} D. R. Taylor and W. J. L. Buyers, Phys. Rev. B \textbf{54}, R3734 (1996).
\bibitem{Fisch} R. Fisch, Phys. Rev. \textbf{51}, 11507 (1995).
\bibitem{Taylor3} D. R. Taylor, E. Zwartz, J.H. Page, 1986, J. Magn. Magn. Mat. \textbf{54-57}, 57 (1986).
\bibitem{It2003} M. Itakura, Phys. Rev. B {\bf 68}, 100405(R) (2003).
\bibitem{Tejada} J. Tejada and X. Zhang, in \textit {Quantum Tunnelling of Magnetization}, edited by L. Gunther and B. Barbara (Kluwer Academic Publishers, Dordrecht, Netherlands, 1995), p. 171.   
\bibitem{BiCaTa2008} O. V. Billoni, S. A. Cannas, F. A. Tamarit, to be published.
\end{thebibliography}

\end{document}